\documentclass[preprint]{elsarticle}

\usepackage{hyperref}
\usepackage{color}

\journal{Exhaled breath barbotage: a new method for pulmonary surfactant dysfunction assessment}

\begin{document}

\begin{frontmatter}

\title{Exhaled breath barbotage: a new method of pulmonary surfactant dysfunction assessing}



\author[mymainaddress]{Aleksey~Mizev}

\author[mymainaddress]{Anastasia~Shmyrova}

\author[mymainaddress]{Irina~Mizeva\corref{mycorrespondingauthor}}
\cortext[mycorrespondingauthor]{Corresponding author}
\ead{mizeva@icmm.ru}

\author[mysecondaryaddress]{Irina~Peleneva}

\address[mymainaddress]{ 614013, Institute of Continuous Media Mechanics,  Korolyov 1, Perm, Russia}
\address[mysecondaryaddress]{614000, Perm State Medical University, Petropavlovskaya st. 26, Perm, Russia}

\begin{abstract}

Exhaled air contains aerosol of submicron droplets of the alveolar lining fluid (ALF), which are generated in the small airways of a human lung. Since the exhaled particles are micro-samples of the ALF, their trapping opens up an opportunity to collect non-invasively a native material from respiratory tract. Recent studies of the particle characteristics (such as size distribution, concentration and composition) in healthy and diseased subjects performed under various conditions have demonstrated a high potential of the analysis of exhaled aerosol droplets for identifying and monitoring pathological processes in the ALF. In this paper we present a new method for sampling of aerosol particles during the exhaled breath barbotage (EBB) through liquid. The barbotage procedure results in accumulation of the pulmonary surfactant, being the main component of ALF,   on the liquid surface, which makes possible the study its surface properties. We also propose a data processing algorithm to evaluate the surface pressure ($\pi$) -- surface concentration ($\Gamma$) isotherm from the raw data measured in a Langmuir trough. Finally, we analyze the $(\pi-\Gamma)$ isotherms obtained for the samples collected in the groups of healthy volunteers and patients with pulmonary tuberculosis and compare them with the isotherm measured for the artificial pulmonary surfactant.
\end{abstract}


\end{frontmatter}

\section{Introduction}
\label{intro}

With every single exhalation a human lung emits an aerosol, which contains small droplets of alveolar lining fluid (ALF) \cite{Fairchild1987,Papineni1997,Fritter1991,Holmgren2010,Schwarz2010,Schwarz2014}. Although the mechanisms responsible for droplet formation are still unclear, the most probable one is associated with the processes of closure and reopening of the airways during normal breathing \cite{Haslbeck2010,Johnson2009}. At the end of exhalation the fluid layer, lining bronchioles, can collapse due to the Rayleigh instability, which results in plug formation and airway closure. During the subsequent inhalation the rupture of the fluid film is accompanied by droplet formation. The studies of the concentration and size distribution of aerosols \cite{Fairchild1987,Papineni1997,Fritter1991,Holmgren2010,Schwarz2010,Schwarz2014} have demonstrated that an exhaled air of a healthy human during normal breathing contains, on average, a few submicron particles per cubic centimeter. In spite of the high reproducibility of measurements within subject, both characteristics show a high inter-subject variability and a strong dependence on the type of breathing and the breathing maneuvers before sampling \cite{Schwarz2010,Schwarz2014}.

Over the last years, the study of the exhaled aerosols has attracted increasing interest as the particles emitted represent micro-samples of the ALF. The analysis of the droplet composition has shown the presence of all components of ALF in undiluted concentration \cite{Olin2013,Tinglev2016,Ullah2015}. From this point of view, the trapping of exhaled particles provides  a new non-invasive way to obtain the native material from the respiratory tract. Recently some studies of the ALF samples obtained with various capturing systems or based on the analysis of separate droplets have been published. The differences in both droplet composition and size distribution in the groups of subjects with chronic obstructive pulmonary disease \cite{Schwarz2014}, asthma \cite{Schwarz2014,Olin2009}, cystic fibrosis \cite{Olin2009} or pulmonary tuberculosis \cite{Wurie2016} in comparison with the healthy ones were reported. These findings indicate a high potential of the exhaled aerosol droplet analysis as a tool to identify and monitor of pathological processes in the ALF.

In this paper, we present a new effective way to capture and to accumulate aerosol particles based on the exhaled breath barbotage (EBB) procedure, which allows one to study the surface-active properties of pulmonary surfactant (PS). Being the main component of the ALF, the PS is a surface-active lipoprotein complex produced in a human lung by type II alveolar cells \cite{notter2000lung}. The most important function of PS is to reduce the alveolar surface tension which results in increasing pulmonary compliance and allows the lung to inflate much more easily, reducing thereby the work of breathing. The unique ability of the compressed PS layer to decrease surface tension to very low, near zero, level prevents as well atelectasis (collapse of lung alveoli) at the end of expiration. A variety of pulmonary diseases (such as asthma, pneumonia, adult respiratory distress syndrome, tuberculosis, etc.) are able to cause surfactant deficiency or to change its composition, which reduces of its surface activity, provoking alveolar instability and development of inflammatory processes in the lung \cite{Hohlfeld2002,Baritussio2004,Wright2001,Willson2008,Schwab2009,Raghavendran2011, Chroneos2009,Chimote2005,Hasegawa2003,Wang2008}. Measurement of the surface-active properties of PS is an effective way to monitor the functional state of a lung surfactant system. In our investigation we show that the accumulation of the captured material of the aerosol particles on the saline surface in a Langmuir through makes it possible to examine the surface-active properties of the collected material immediately after the barbotage procedure. The proposed data processing algorithm allows us to obtain the surface pressure ($\pi$) - surface concentration ($\Gamma$) isotherm from the raw data obtained in the experiment. Finally we also analyze the $(\pi-\Gamma)$ isotherms obtained for the samples collected in the groups of healthy volunteers and patients with lung tuberculosis and compare them with the isotherm measured for the artificial pulmonary surfactant.

\section{Materials and methods}

\subsection{Samples collection method}

\emph{ Exhaled breath barbotage}

We use the procedure of barbotage (or bubbling) of exhaled air through liquid to capture  ALF droplets. The complex air motion occurred in the course of bubble formation together with a large area of the liquid-gas interface should lead to more frequent mechanical contacts between the droplets and the bubble surface. Collisions of the droplets with the surface of bubbles initiate their spreading, which facilitates the adsorption of PS. The latter accumulates at the upper liquid-air interface as the bubbles rise. The adsorbed layer can be further investigated by tensiometric methods.

This idea lies behind the design of the collecting system shown in Fig.1. The system consists of a Teflon tube connected to a saliva trap through which a patient breathes out. The other end is immersed in the saline solution filling the Langmuir trough. Since, the inner diameter of the tube is a compromise between two requirements, it should be large enough to help the patients breathe easier during the procedure. On the other hand, the smaller is the diameter of bubbles, the more frequent are the contacts between the droplets and the bubble surface, and the higher is the efficiency of surfactant collection. The inner diameter of the tube is 3~mm.

\begin{figure}
\begin{center}
\resizebox{0.7\columnwidth}{!}{\includegraphics{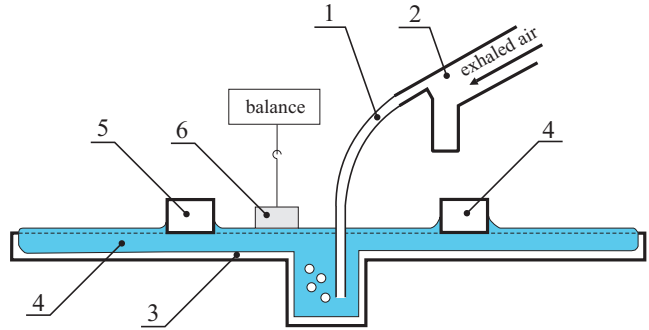}}
\caption{Experimental set \textbf{up} for exhaled breath barbotage and examination of the dynamic surface-active properties of samples: 1 - Teflon tube, 2 - saliva trap, 3 - Teflon trough, 4 - saline subphase, 5 - Delrin barriers, and 6 - platinum Wilhelmy plate.}
\label{fig:1}
\end{center}
\end{figure}

The collected material (saline subphase with PS adsorbed on its surface) will be henceforth called the exhaled breath barbotate (EBB) by analogy with the exhaled breath condensate (EBC) samples.

At the beginning of the procedure, a subject took a deep breath and then breathed completely out through the mouth into the tube. After that, the bubbling was stopped and one compression-expansion circle in the Langmuir trough was realized. After the measurements were done, the subject was asked to exhale again. This procedure was repeated several times, which allowed us to obtain a ($\pi-S$) isotherms after a few exhalations. \\

\noindent \emph{Artificial pulmonary surfactant}

A series of tests was conducted with the artificial surfactant "Surfactant-BL" ("Biosurf", Russia) obtained by the extraction from bovine minced lung. The drug was dissolved in chloroform up to the concentration of $3.36\cdot{{10}^{-5}}g/ml$. A known amount of the solution was spread by a microsyringe on the saline subphase surface in a drop-wise manner and was allowed to elapse for complete solvent evaporation (~10 min).

\subsection{Surface activity study}

The study of the surface activity of the EBB was performed with a KSV Minimicro system (KSV, Finland). A Teflon rectangular trough ($195\times50$ $mm^2$) was filled with saline (Fig.\ref{fig:1}) prepared on the basis of high purity water (with conductivity less than 0.2  $\mu$S/cm) to avoid the influence of impurities. The surface of the subphase was cleaned by gently aspirating, and the cleanliness was confirmed by a zero reading of surface pressure. Measurements were carried out under controled temperature ($37.0\pm0.2$) $^{\circ}C$. The presence of the hollow in the center of the trough allowed us to bubble the exhaled breath directly through the saline subphase filling the trough. Two Delrin barriers were employed to provide symmetric film compression. The same barrier moving speed (5 mm/min per barrier) was used for both film compression and expansion. Expansion started directly after the compression was finished. The surface pressure $\pi$ was continuously monitored by a platinum Wilhelmy plate during barrier moving. The ($\pi-S$) isotherm was measured during one compression-expansion cycle. In order to study the accumulation of the pulmonary surfactant on the saline surface, we measured such an isotherm after each of several exhalations of every subject.

\subsection{Groups of subjects}

Two groups of subjects were involved in the study. The control group comprised 15 healthy nonsmoking volunteers (7 males and 8 females) aged 23 to 43 years. The second group included 20 tubercular (TB) patients (13 males and 7 females) aged 18 to 48 years. Table \ref{Tab1} summarizes the main characteristics of the groups. In preliminary experiments, it was found that the results obtained from the same patient are significantly dependent on the patient's motion activity before the barbotage procedure. In order to provide for the same initial conditions of every test, all subjects were asked to make a few exercises (e.g. to perform several squats) just before the beginning of the procedure. This allowed us to increase essentially the intra-subject reproducibility of the results.

The Ethical Committee of the Perm State Medical University approved the study protocol, and in all cases informed an consent was received from the patients.

\begin{table}[h]
\caption{\label{jlab1} The main characteristics of the groups under study}
\begin{tabular}{c  c c }
\\\hline
&Control &TB group \\
& n=15 & n=20 \\\hline
Sex (M/F) & 7/8 & 13/7 \\
Age, years & $28\pm4$ & $33\pm9$ \\
Infiltrative pulmanory TB in disintegration phase & -- & 13 \\
Fibrous-cavernous pulmonary TB & -- & 2 \\
Disseminated pulmonary TB & & \\
in infiltration and disintegration phases & -- & 5 \\ \hline
\end{tabular}
\label{Tab1}
\end{table}

%\begin{table}
%\caption{\label{jlab1} The main characteristics of the groups under study} \footnotesize
%\begin{center}
%\begin{tabular}{@{}p{7cm}cc}
%&Control (N=15)&TB group (N=20)\\
%Sex (M/F) & 7/8 & 13/7 \\
%Age, years & $\left( 28\pm 4 \right)$ & $\left( 33\pm 9 \right)$ \\
%Infiltrative pulmanory TB in disintegration phase & -- & 13 \\
%Fibrous-cavernous pulmonary TB & -- & 2 \\
%Disseminated pulmonary TB in infiltration and disintegration phases & -- & 5 \\
%\end{tabular}\\
%\end{center}
%\end{table}
%\normalsize

\subsection{Data analysis}
\label{sec:dataproc}

\begin{figure}
\begin{center}
\resizebox{0.52\columnwidth}{!}{\includegraphics{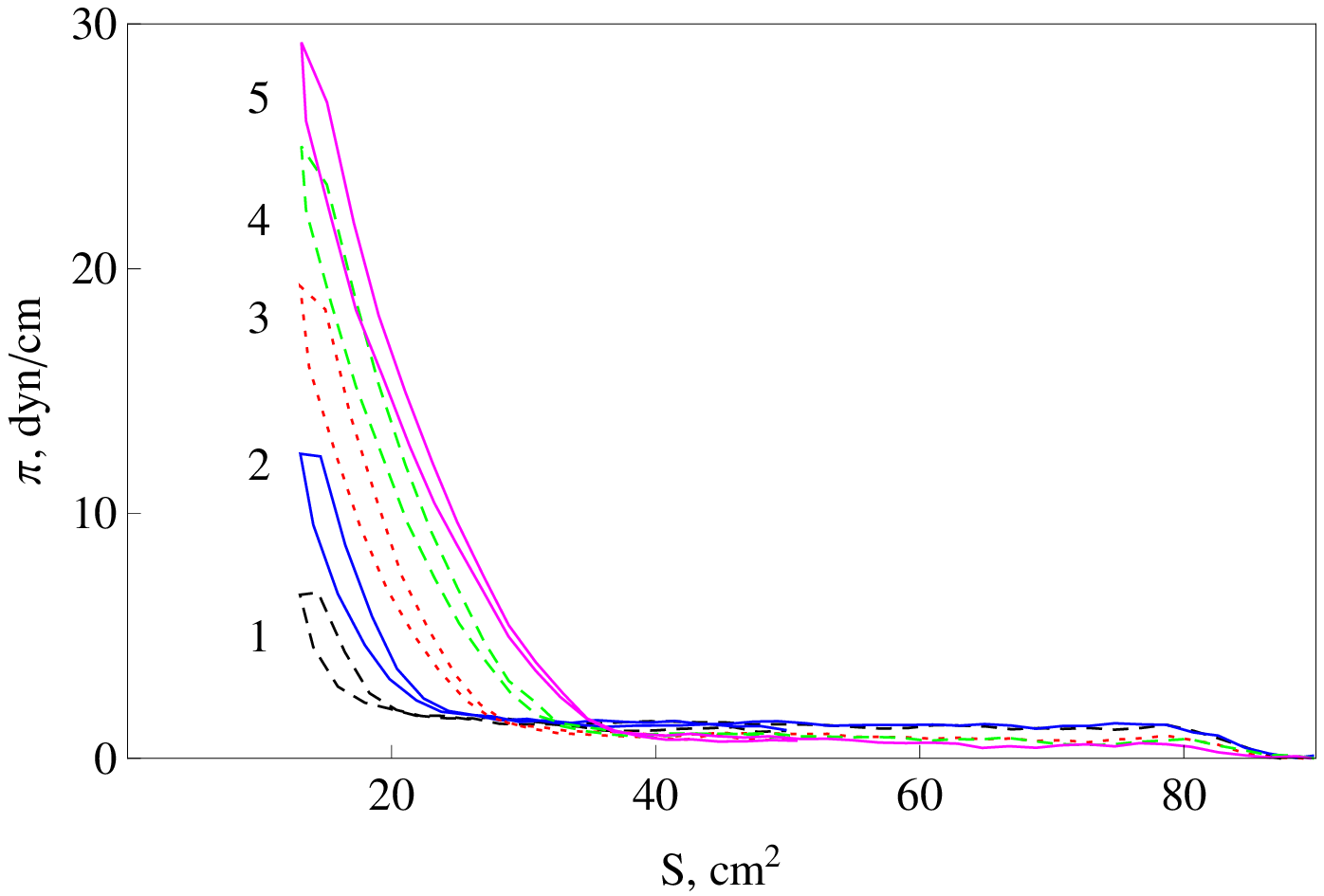}} \resizebox{0.2\columnwidth}{!}{\includegraphics{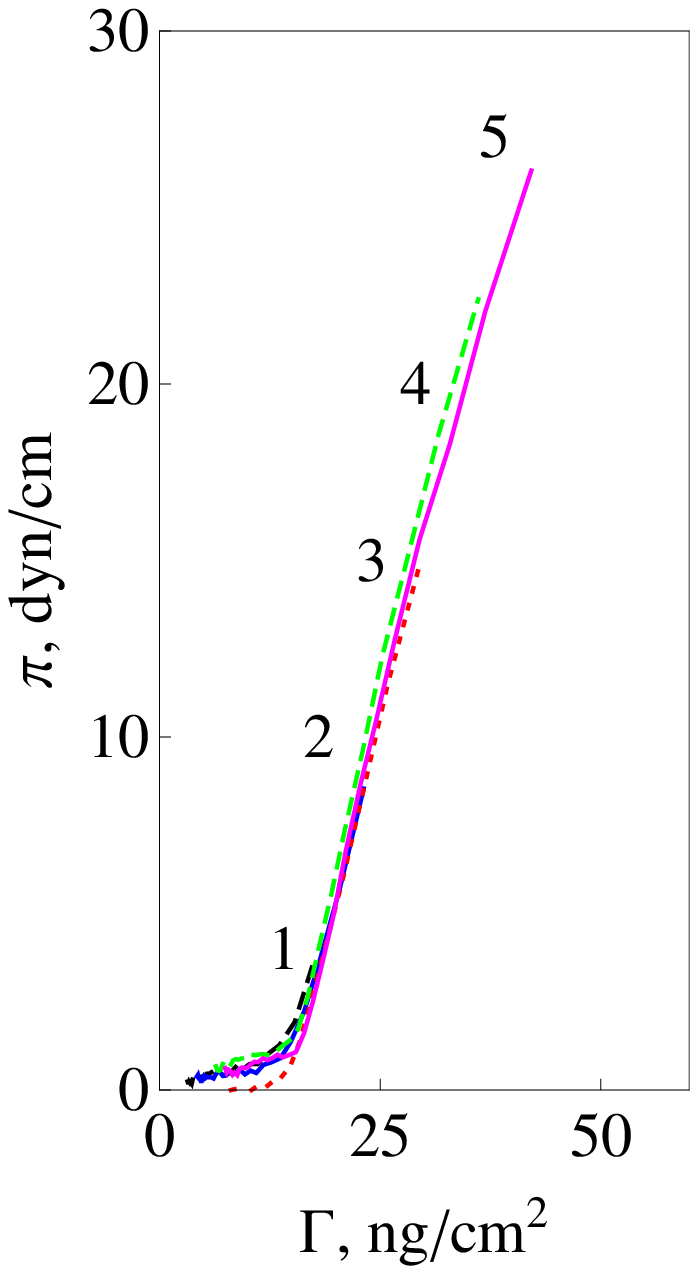}}
\resizebox{0.203\columnwidth}{!}{\includegraphics{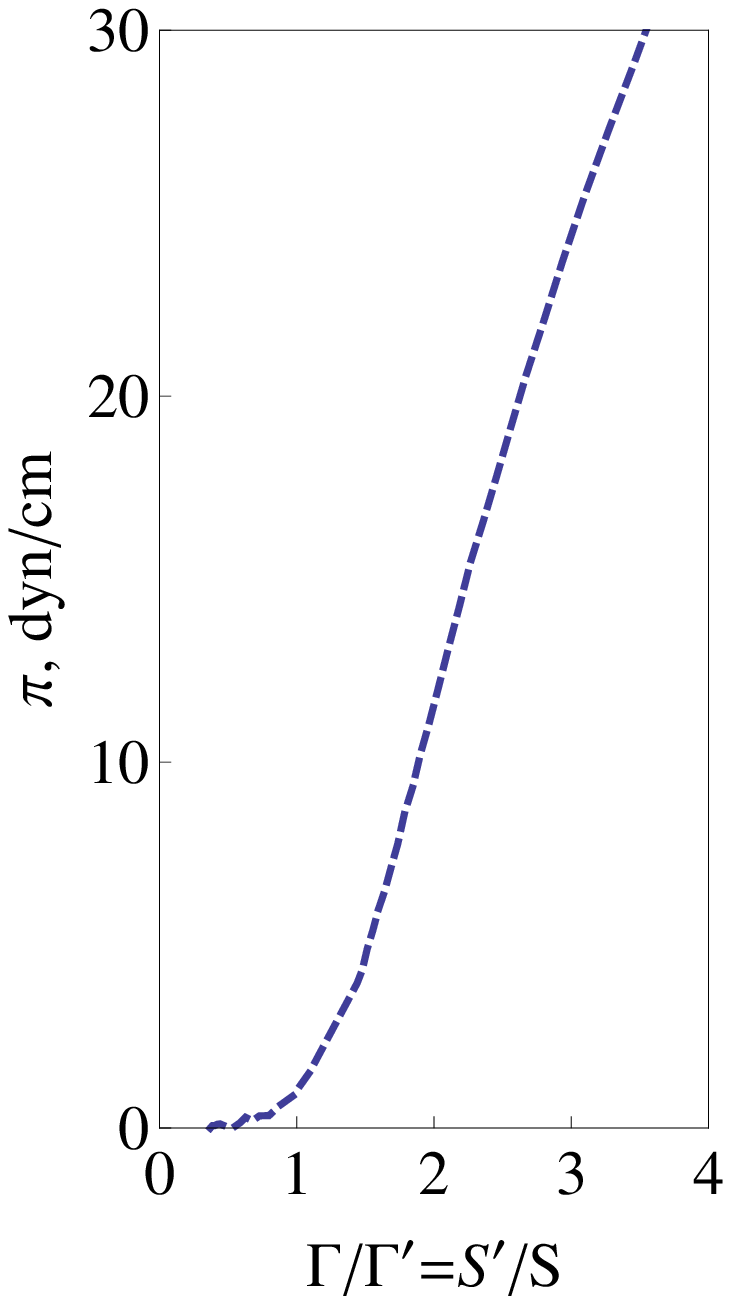}}
\caption{Graphes illustrating different steps of the data analysis. (a) compression-expansion isotherms obtained for the artificial surfactant of different initial surface concentration: 1 - 11 $ng\cdot {{ cm}^{-2}}$, 2 - 15 $ng\cdot {{ cm}^{-2}}$, 3 - 17 $ng\cdot {{ cm}^{-2}}$, 4 - 20 $ng\cdot {{ cm}^{-2}}$, 5 21 $ng\cdot {{ cm}^{-2}}$; (b) $(\pi-\Gamma)$ isotherm; (c) $(\pi-\Gamma)$ isotherm measured in terms of $\Gamma^\prime$.}
\label{fig:2}
\end{center}
\end{figure}

The efficiency of the data processing algorithm is demonstrated using the experimental data obtained for the artificial surfactant. The $(\pi-S)$ isotherms measured for five different initial surface concentrations of the artificial surfactant are presented in Fig.\ref{fig:2}a. Compression and expansion curves do not coincide due to the hysteresis phenomenon. Further we consider only the compression part of each circle. The maximum surface pressure $\pi_{max}$ corresponding to the minimal $S$ increases with each addition of the surfactant to the saline surface. This parameter can be used for  quantification of the surfactant content on the subphase surface. Each curve consists of two parts differing in slope, which results from the phase transition in the surface layer from gas (low slope part) to liquid-expanded (high slope part) phase state. This phase transition takes place at the surface concentration $\Gamma^\prime$, the value of which is the internal property of the surfactant. Since the curves in Fig.\ref{fig:2}a are the fragments of the same $(\pi-\Gamma)$ isotherm, we can reconstruct it from the raw data by calculating the surface concentration.

\begin{equation}
\Gamma=\Gamma_0 \frac{S_0}{S},
\end{equation}

\noindent where $\Gamma_0$  is the known initial surface concentration which is different for each curve. The calculated results is shown in Fig.\ref{fig:2}b. All curves coincide, forming the  $(\pi-\Gamma)$ isotherm, which is the main characteristic of any surfactant.

The major problem during the study of EBB samples in the Langmuir trough is the inability to measure the initial surface concentration of the surfactant adsorbed at the interface during breathing. This fact does not allow the reconstruction of the $(\pi-\Gamma)$ isotherm from the $\pi(S)$ raw data. On the other hand, the required dependence can be found in relative units. For any surface concentration $\Gamma$, we can write the following expression:

\begin{equation}
\Gamma / \Gamma^\prime= S^\prime/S,
\label{Eq.Norm}
\end{equation}

\noindent where $S$ and $S^\prime$ are the current surface area and the surface area corresponding to phase transition, which are directly measured in the experiment. By changing the coordinates from $\pi(S)$ to $\pi(S^\prime/S)$, we obtain the required isotherm, where the surfactant surface concentration is measured in terms of $\Gamma^\prime$. Such a normalized isotherm for the artificial surfactant is given in Fig.\ref{fig:2}c.

During the EBB procedure, each exhalation adds the unknown amount of pulmonary surfactant to the saline surface. Since a certain subject emits the same surfactant during each exhalation, we apply normalization (\ref{Eq.Norm}) to the EBB sample data. To determine $S^\prime$, every $\pi(S)$ curve was fitted by two linear functions. The abscissa of their intersection was assumed as $S^\prime$. Hence, we get the non-dimensional parameter $S^\prime/S$, which describes the normalized surface concentration of  the surfactant. For statistical analysis, we took the surface pressure $\pi$ at different $S^\prime/S$ and compared the results of the groups. The maximum surface pressure ($\pi_{max}$) as a function of total exhalation time $t$ was used as a measure of the pulmonary surfactant accumulation rate during the EBB procedure for each subject. To this end, we interpolated the dependence $\pi_{max}(t)$ by a piecewise linear function to detect ($\pi_{max}$) at the 20th, 40th, 60th and 80th seconds.

The results are represented as $median\pm standart deviation$. For unpaired data,  differences between the groups were evaluated using the Mann-Whitney criterion (p$<$0.05 was assumed to be statistically significant). In the statistical analysis of the measured values, we did not assume that they had normal distribution;  that is why a nonparametric statistical test was carried out. Data preprocessing and statistical analysis were performed using Mathematica 8.0.

\section{Results}

First, the intra-reproducibility of the EBB technique was estimated. For this purpose, the EBB samples of one person (female, aged 28 years) were collected during 5 days (1 sample a day) under  similar conditions. We used the data processing technique described above. It has been found that the variability of $\pi_{max}(t)$ and $\pi(S^\prime/S)$ is low and the divergence is less than 10\%, which indicates rather good reproducibility of the EBB method.

\begin{figure}
\begin{center}
\resizebox{0.4\columnwidth}{!}{\includegraphics{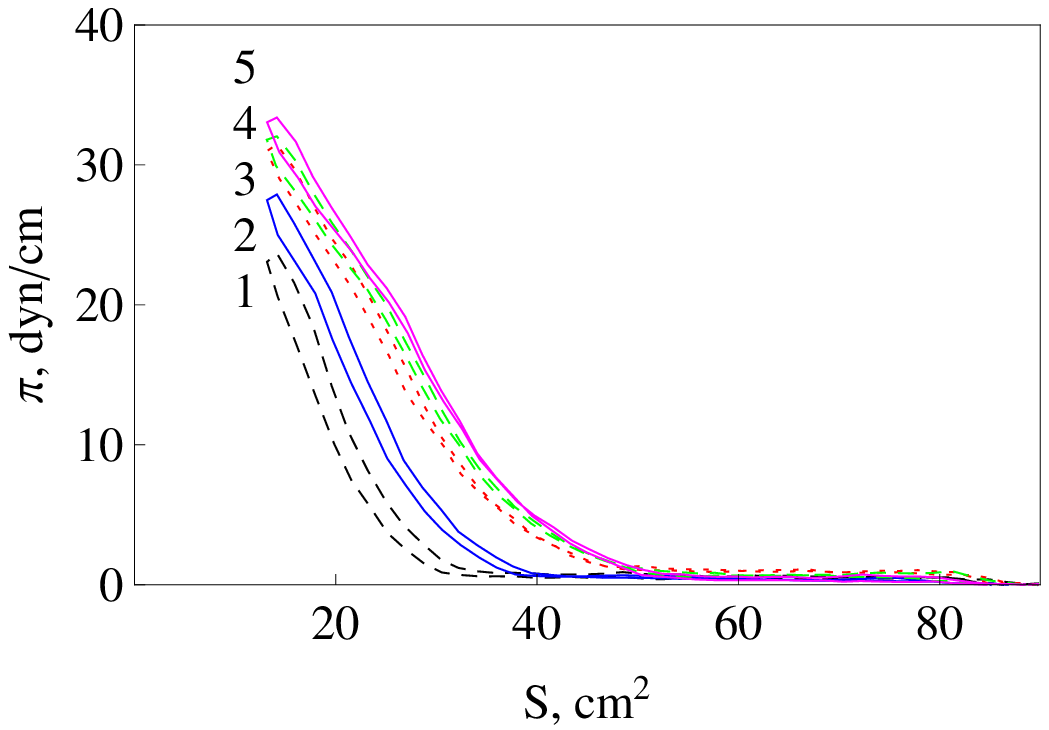}} \resizebox{0.4\columnwidth}{!}{\includegraphics{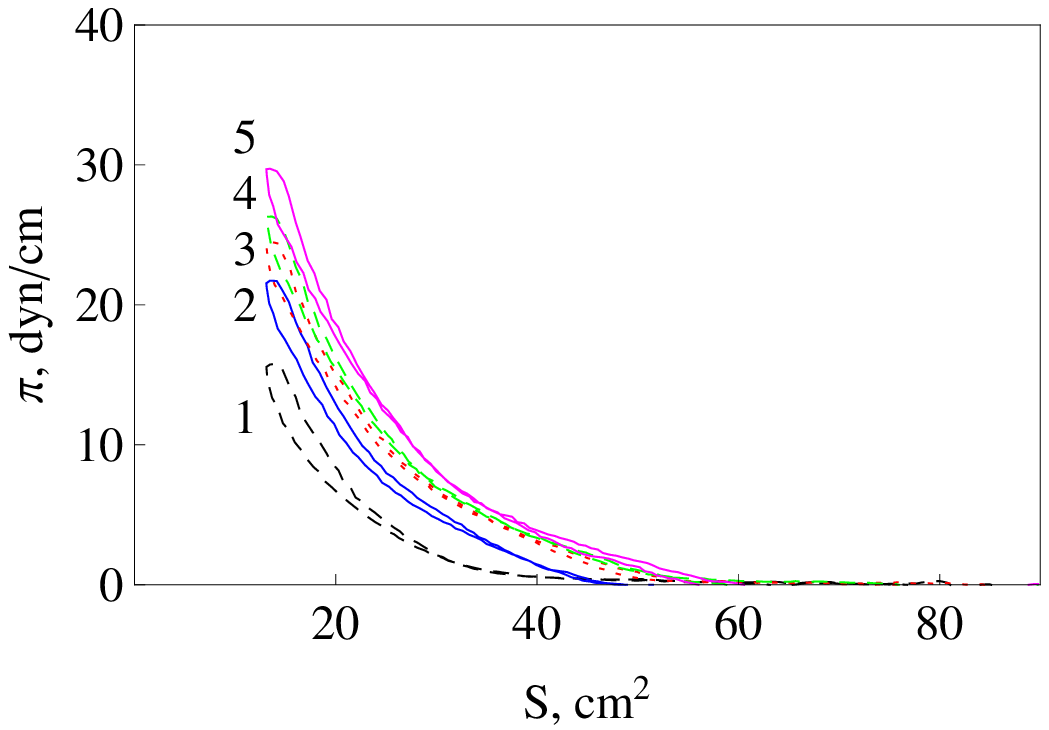}}
\resizebox{0.4\columnwidth}{!}{\includegraphics{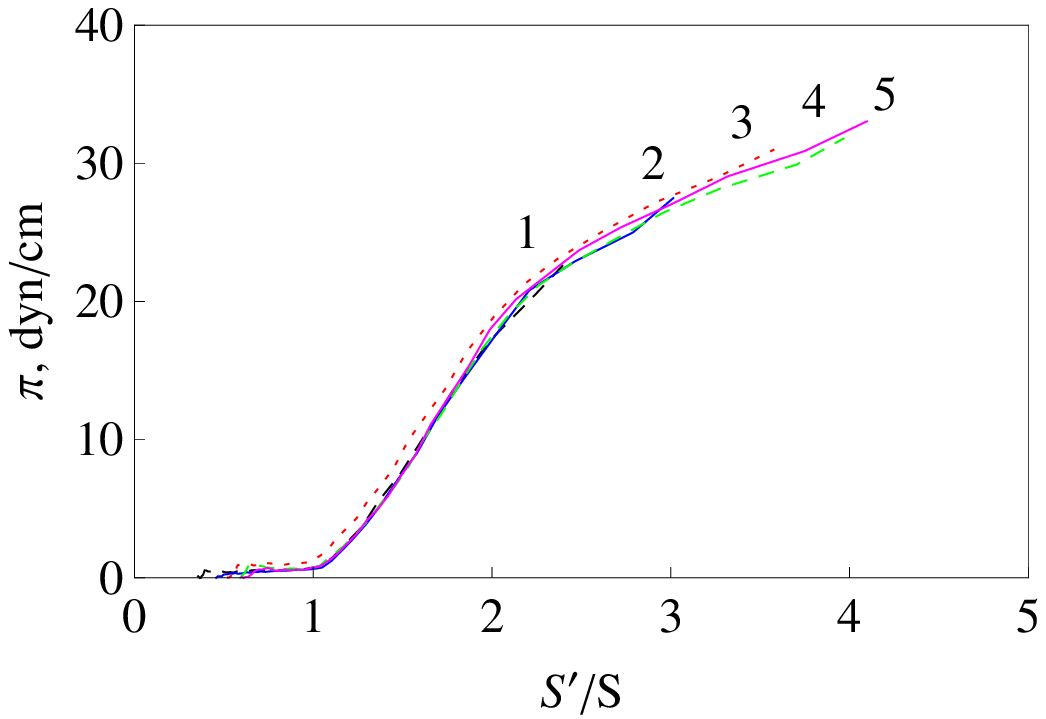}} \resizebox{0.4\columnwidth}{!}{\includegraphics{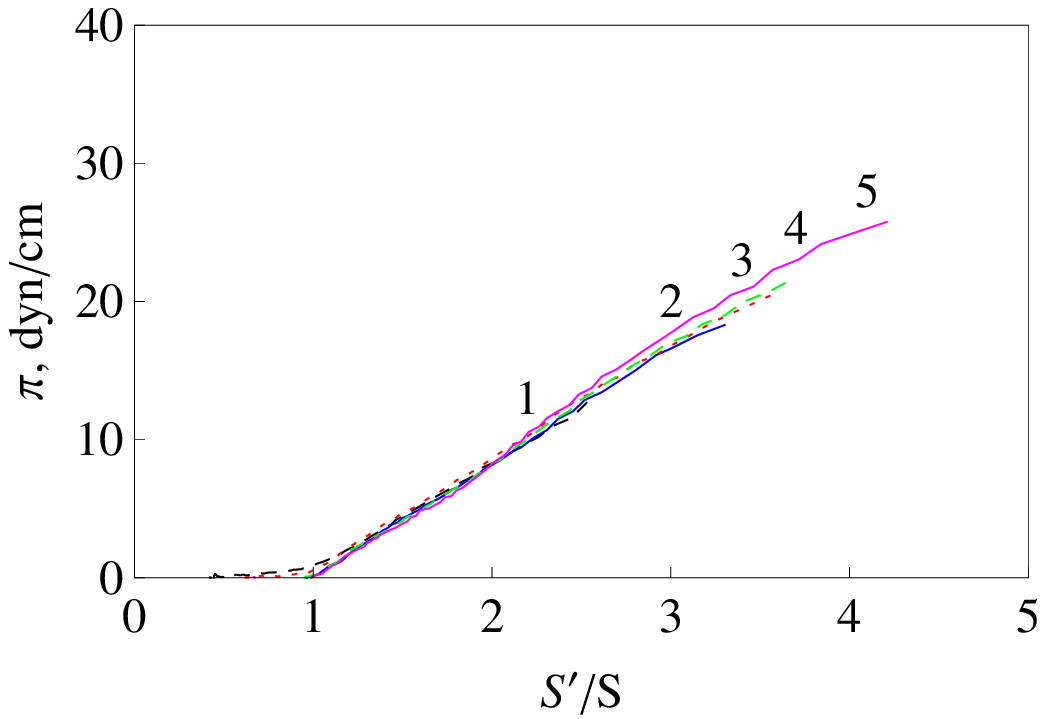}}
\caption{$(\pi-S)$ and normalized $\pi(S^\prime/S)$ isotherms obtained after each of five exhalations of a healthy volunteer (a and c) and a TB patient (b and d).}
\label{fig:3}
\end{center}
\end{figure}

The $(\pi-S)$ isotherms for the compression-expansion circle, measured after each of five exhalations of a healthy volunteer, are presented in Fig.\ref{fig:3}a. In the further analysis, we consider only the compression part of each circle. Each subsequent curve lies above the previous one and $\pi_{max}$ increases after each exhalation, which indicates the accumulation of PS on the saline surface during the EBB procedure. A comparison of the results presented in Fig.\ref{fig:2}a and Fig.\ref{fig:3}a allows us to estimate the average amount of the PS collected by the EBB method in healthy subjects. It makes up approximately 0.7 $\mu g$ per exhalation.

\begin{figure}
\begin{center}
\resizebox{0.7\columnwidth}{!}{\includegraphics{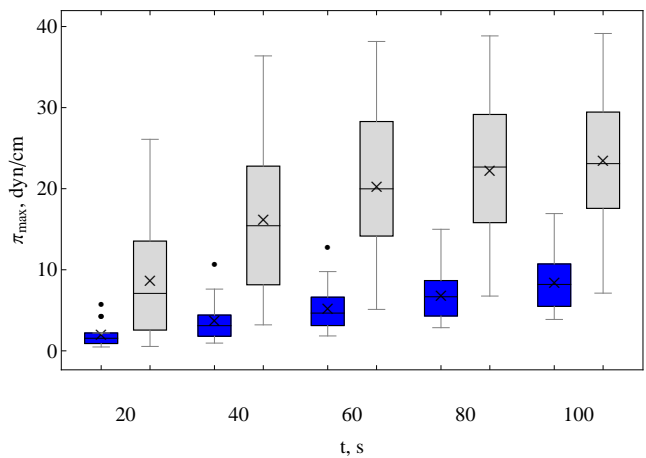}}
\caption{Maximum surface pressure $\pi_{max}$ obtained in the groups of healthy volunteers (shaded boxes) and TB patients (open boxes) after a certain time. All differences are significant (p$<$0.05)} \label{fig:4}
\end{center}
\end{figure}

The $(\pi-S)$ isotherms obtained for the EBB samples of one TB patient are presented in Fig.\ref{fig:3}b. The curves look like those measured for the healthy volunteer (Fig.\ref{fig:3}a). However if we take into account the shorter, on average, exhalation time of TB patients, then we find a higher rate of PS accumulation at the interface for this group. To summarize the results for all subjects, we evaluated the distribution of $\pi_{max}$ at the same moments of the total exhalation time for control group and TB patients (Fig.\ref{fig:4}). Note that $\pi_{max}$ grows faster with time for the TB patients, which indicates the higher rate of emission of aerosol particles during breathing.

The $\pi_{max}(t)$ dependence allows us to study the variations of the aerosol particle concentration in the exhaled air. To quantify the variations of PS properties, one needs to compare the normalized $(\pi-\Gamma / \Gamma^\prime)$ isotherms. The results of applying the data processing algorithm described in \textbf{Sec.\ref{sec:dataproc}} to the raw data given in (Fig.\ref{fig:3}a) and (Fig.\ref{fig:3}b) are presented in Fig.\ref{fig:3}c and Fig.\ref{fig:3}d for a healthy volunteer and a TB patient, respectively. In is seen that all curves form a single dependence in the $\pi(S^\prime/S)$ plane. Each subsequent curve corresponding to the subsequent exhalation extends to higher surface concentrations in the $(\pi-\Gamma / \Gamma^\prime)$ isotherm, which reflects the accumulation process. It is obvious that the first several exhalations are necessary to accumulate the surfactant sufficient to obtain the required isotherm. Thus, there is no need to measure the $\pi-S$ isotherm after every exhalation; it suffice to process the data obtained after the last exhalation.

\begin{figure}
\begin{center}
\resizebox{0.7\columnwidth}{!}{\includegraphics{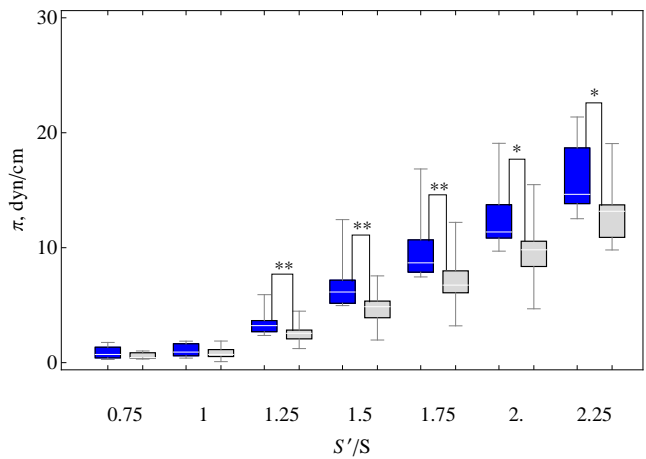}}
\caption{Box-and-Whisker diagram for the surface pressure obtained from the normalized isotherms $\pi(\Gamma / \Gamma^\prime)$; blue rectangles correspond to controls and the gray ones to TB patients. Stars indicate the level of significance estimated by Mann-Whitney criteria: $\ast$- p$<$0.05, $\ast\ast$-p$<$0.01.}
\label{fig:5}
\end{center}
\end{figure}

In Fig.\ref{fig:5}, we generalize all normalized isotherms for the control and TB groups. To this end, $\pi(\Gamma / \Gamma^\prime)$ for $\Gamma / \Gamma^\prime$={0.75,1,1.25,1.5,1.75,2} were calculated for all subjects and the results were summarized using Box-and-Whisker plots. The difference between the healthy and TB subjects is not significant in the range of $\Gamma / \Gamma^\prime < 1$ when the surfactant layer is in a gaseous phase state. The difference becomes significant (p$<$0.05) at larger $\Gamma / \Gamma^\prime$. Surface pressure $\pi(\Gamma/\Gamma^\prime)$ values are summarized in Table\ref{Tab2}.

For comparison we analyze the results obtained for the artificial surfactant. The surface pressure values for healthy subjects coincide with those for the artificial surfactant at the defined level of significance ($p=0.05$). On the contrary, the isotherms of TB subjects demonstrate significant difference from the isotherms of both the artificial system and the control group. The surface pressure values are smaller in TB group which says of the less surface activity of the samples.

%\begin{table}[h]
%\caption{Sodium parameters} \label{tab:av1}
%\begin{tabular}{|c|c|}
%\hline
%Name of the parameter      & Value                \\ \hline
%\end{tabular}
%\end{table}

\begin{table}[h]
\caption{ The surface pressure $\pi$ under different values of surface density $\Gamma / \Gamma^\prime$ for artificial PS and EBB samples of control and TB groups}
\begin{tabular}{c c c c }
\\hline
& Surface pressure &$\pi, dyn/cm$\\
$\Gamma / \Gamma^\prime$ & artificial         & CG                & TB             \\ \hline
   1.0          & $0.99$        &  $1.0\pm0.5$, NS  & $0.8\pm0.5$, NS          \\
   1.25          & $3.3$         &  $3.3\pm0.9$, NS  & $2.4\pm0.9$, p=0.005          \\
   1.5           & $6.8$         &  $6.5\pm2.0$, NS  & $4.5\pm1.4$, p=0.0001          \\
   1.75          & $10.4$        &  $10.1\pm2.7$, NS & $7.2\pm2.2$, p=0.0002          \\
   2.0           & $14.0$        &  $13.0\pm2.9$, NS & $10.1\pm3.0$, p=0.008          \\
   2.25          & $17.3$        &  $16.2\pm3.2$, NS  & $13.5\pm3.0$, p=0.009          \\
   2.5           & $20.5$        &  $18.6\pm3.2$, NS  & $16.7\pm2.9$, p=0.014          \\
   2.75          & $23.4$        &  $22.4\pm4.1$, NS  & $19.2\pm3.1$, p=0.003          \\ \hline
\end{tabular}
\label{Tab2}
\end{table}

%\begin{table}[h]
%\caption{Statistics for $\pi, dyn/cm$} \label{tab:av1}
%\begin{tabular}{|c|c|c|c|}
%\hline $\Gamma / \Gamma^\prime$   & artificial         & CG                & TB             \\ \hline
%    1.0          & $0.99$        &  $1.0\pm0.5$, NS  & $0.8\pm0.5$, NS          \\
%   1.25          & $3.3$         &  $3.3\pm0.9$, NS  & $2.4\pm0.9$, p=0.005          \\
%   1.5           & $6.8$         &  $6.5\pm2.0$, NS  & $4.5\pm1.4$, p=0.0001          \\
%   1.75          & $10.4$        &  $10.1\pm2.7$, NS & $7.2\pm2.2$, p=0.0002          \\
%   2.0           & $14.0$        &  $13.0\pm2.9$, NS & $10.1\pm3.0$, p=0.008          \\
%   2.25          & $17.3$        &  $16.2\pm3.2$, NS  & $13.5\pm3.0$, p=0.009          \\
%   2.5           & $20.5$        &  $18.6\pm3.2$, NS  & $16.7\pm2.9$, p=0.014          \\
%   2.75          & $23.4$        &  $22.4\pm4.1$, NS  & $19.2\pm3.1$, p=0.003          \\\hline
%\end{tabular}
%\label{Tab2}
%\end{table}

\section{Discussion}

In this paper we have demonstrated the possibility of examining the surface active properties of the PS obtained by the new noninvasive way, which is based on capturing the aerosolized ALF droplets in the course of exhaled breath barbotage. Since the surfactant is accumulated on the surface of the saline subphase filling the Langmuir trough, it becomes possible to measure the surface properties immediately after collecting, escaping the preliminary extraction or separation procedures, as it takes place for EBC or BAL. The proposed technique is simple and inexpensive to implement and effortless to patients which makes it preferable for kids, elder people or patients with severe diseases. The EBB surface properties demonstrate high reproducibility in both intra-subject and inter-subject tests. The data processing allowing one to reconstruct the $(\pi-\Gamma)$ isotherm in terms of $\Gamma^\prime$ can be useful for studies where it is impossible to measure surfactant concentration directly.

It is interesting to discuss the sensitivity of the method to the rate of aerosol particle emission. Recently it has been shown \cite{Schwarz2010,Schwarz2014} that the variation of particle emission is low within subjects and it can be rather high between subjects. In our study, the higher concentration of the particles in expired air would result in the higher concentration of the pulmonary surfactant on the saline surface at the end of one exhalation and, therefore, the larger value of $\pi_{max}$ at the end of surface compression. Thus, the rate of $\pi_{max}$ growth with total exhalation time provides the information about the emission rate. On the other hand, the emission rate in no way can influence the $\pi(\Gamma / \Gamma^\prime)$ isotherm. Actually the form of this isotherm is defined by the surface activity of PS rather than by its amount. The low emission rate just means that a subject should make more exhalations to reach the higher surface concentration values in the isotherm. Thus, the $\pi(\Gamma / \Gamma^\prime)$ isotherm is insensitive to the particle concentration in expired air and can be used in a comparative analysis of the results obtained for different subjects independently on the emission rate of particles.

These conclusions are confirmed by the dispersion of the data presented in Fig.\ref{fig:4} and Fig.\ref{fig:5}. Rather high variability of $\pi_{max}$ ($SD/M\simeq 60\%$) inside both groups reflects the existing differences in the emission rate between the subjects. At the same time, the results presented in $\pi(\Gamma / \Gamma^\prime)$ isotherms (Fig.\ref{fig:5}) demonstrate relatively high inter-individual repeatability in both groups ($SD/M\simeq 30\%$ for control group and $SD/M\simeq 40\%$ for TB patients), which verifies the isotherm form independence of the concentration of ALF particles in expired air.

Another positive aspect of the proposed method consists in the fact that the EBB technique eliminates the need for the control group of healthy subjects. The $(\pi-\Gamma / \Gamma^\prime)$ isotherm measured for artificial surfactant with well controlled content can be applied as a standard for calibration purposes. Moreover, the comparison with the $\pi(\Gamma / \Gamma^\prime)$ isotherms measured for artificial mixtures containing not all components of the pulmonary surfactant complex can be used for better interpretation of the abnormalities obtained in the $\pi(\Gamma / \Gamma^\prime)$ isotherms of subjects.

Let us now discuss the results obtained in the control and TB groups. First, an essential difference in the growth rate of $\pi_{max}$ during the total exhalation time (Fig.\ref{fig:4}) should be noted. The main mechanism of ALF droplets generation is the rupture of the plugs formed in airways due to the Rayleigh instability of the liquid lining \cite{Haslbeck2010,Johnson2009}. Reopening of an airway during breathing is accompanied by droplets formation. As was shown in \cite{Grotberg2011,Halpern1993}, an increase in the relation between the liquid layer thickness and the airway diameter makes the liquid lining less stable, which provokes more frequent formation of plugs. The inflammatory processes taking place in the lung in a case of TB result in the narrowing of the diameter of distal airways. This leads to a higher level of the particle emission rate, which results in the higher growth rate of $\pi_{max}$ with total exhalation time observed in the TB group. Similar increase in the particle emission rate in the patients with pulmonary TB was earlier found in experiments, where the exhaled particles were directly counted by optical methods \cite{Wurie2016}.

Secondly, the results measured in the groups differ in the form of the $(\pi-\Gamma / \Gamma^\prime)$ isotherm (Fig.\ref{fig:5}). The isotherms obtained in the TB group have smaller inclination, which indicates the lower surface activity of the EBB samples collected in the TB patients. This fact is in agreement with the previous studies \cite{Wang2008,Chimote2005,Chimote2008}, where the influence of pulmonary TB on the biophysical properties of pulmonary surfactant was investigated. It has been found that the peripheral wall lipids of Mycobacterium tuberculosis reduce the surface activity of the pulmonary surfactant complex. At the same time, it is known \cite{Grotberg2011,Halpern1993} that the lower surface activity leads to a less stable liquid lining in airways. This fact correlates with a higher emission rate in the TB group discussed above.

Finally, we would like to discuss the possible applications of the proposed method, EBB sampling and examination of the collected material, in medical practice. The method is difficult to use for diagnostic purposes because different lung diseases can result in the pulmonary surfactant abnormalities. Being a non-invasive and simple way of assessing the lung surfactant functional state the method can be useful both for screening testing and treatment monitoring.

\section*{Acknowledgment}

The work is supported by the Russian foundation for basic research RFBR-ra under project 17-41-590095

%\bibliography{Barb}

%\begin{thebibliography}{9}
%\bibitem{111}
%Gunawardena H., Harris N.D., Carmichael C. et al. Maximum blood flow and microvascular regulatory responses in systemic sclerosis Rheumatology (Oxford) 2007;46:1079-82.
%\end{thebibliography}

%\bibliographystyle{elsarticle-harv}
%\bibliography{ref_Micro1}

\begin{thebibliography}{10}
\expandafter\ifx\csname url\endcsname\relax
  \def\url#1{\texttt{#1}}\fi
\expandafter\ifx\csname urlprefix\endcsname\relax\def\urlprefix{URL }\fi
\expandafter\ifx\csname href\endcsname\relax
  \def\href#1#2{#2} \def\path#1{#1}\fi

\bibitem{Fairchild1987}
C.~Fairchild, J.~Stampfer, Particle concentration in exhaled breath, American
  Industrial Hygiene Association Journal 48~(11) (1987) 948--949.

\bibitem{Papineni1997}
R.~Papineni, F.~Rosenthal, The size distribution of droplets in the exhaled
  breath of healthy human subjects, Journal of Aerosol Medicine: Deposition,
  Clearance, and Effects in the Lung 10~(2) (1997) 105--116.

\bibitem{Fritter1991}
D.~Fritter, et~al., Experiments and simulation of the growth of droplets on a
  surface (breath figures), Physical Review A 43~(6) (1991) 2858--2869.

\bibitem{Holmgren2010}
H.~Holmgren, E.~Ljungstrom, A.-C. Almstrand, B.~Bake, A.-C. Olin, Size
  distribution of exhaled particles in the range from 0.01 to 2.0 $\mu$m,
  Journal of Aerosol Science 41~(5) (2010) 439 -- 446.

\bibitem{Schwarz2010}
K.~Schwarz, H.~Biller, H.~Windt, W.~Koch, J.~M. Hohlfeld, {Characterization of
  Exhaled Particles from the Healthy Human Lung-A Systematic Analysis in
  Relation to Pulmonary Function Variables}, {Journal of Aerosol Medicine and
  Pulmonary Drug Delivery} {23}~({6}) ({2010}) {371--379}.

\bibitem{Schwarz2014}
K.~Schwarz, H.~Biller, H.~Windt, W.~Koch, J.~M. Hohlfeld, {Characterization of
  Exhaled Particles from the Human Lungs in Airway Obstruction}, {Journal of
  Aerosol Medicine and Pulmonary Drug Delivery} {28}~({1}) ({2015}) {52--58}.

\bibitem{Haslbeck2010}
K.~Haslbeck, K.~Schwarz, J.~M. Hohlfeld, J.~R. Seume, W.~Koch, Submicron
  droplet formation in the human lung, Journal of Aerosol Science 41~(5) (2010)
  429 -- 438.

\bibitem{Johnson2009}
G.~R. Johnson, L.~Morawska, {The Mechanism of Breath Aerosol Formation},
  {Journal of Aerosol Medicine and Pulmonary Drug Delivery} {22}~({3}) ({2009})
  {229--237}.

\bibitem{Olin2013}
A.-C. Olin, {Particles in Exhaled Air-A Novel Method of Sampling Non-Volatiles
  in Exhaled Air}, in: {Amann, A and Smith, D} (Ed.), {Volatile Biomarkers:
  Non-invasive Diagnosis in Physiology and Medicine}, {2013}, pp. {383--391}.

\bibitem{Tinglev2016}
A.~D. Tinglev, S.~Ullah, G.~Ljungkvist, E.~Viklund, A.-C. Olin, O.~Beck,
  {Characterization of exhaled breath particles collected by an electret filter
  technique}, {Journal of Breath Research} {10}~({2}).

\bibitem{Ullah2015}
S.~Ullah, S.~Sandqvist, O.~Beck, {Measurement of Lung Phosphatidylcholines in
  Exhaled Breath Particles by a Convenient Collection Procedure}, {Analitical
  Chemistry} {87}~({22}) ({2015}) {11553--11560}.

\bibitem{Olin2009}
A.~Almstrand, E.~Ljungstrom, J.~Lausmaa, B.~Bake, P.~Sjovall, A.~Olin, {Airway
  Monitoring by Collection and Mass Spectrometric Analysis of Exhaled
  Particles}, {Analitical Chemistry} {81}~({2}) ({2009}) {662--668}.

\bibitem{Wurie2016}
F.~B. Wurie, S.~D. Lawn, H.~Booth, P.~Sonnenberg, A.~C. Hayward, {Bioaerosol
  production by patients with tuberculosis during normal tidal breathing:
  implications for transmission risk}, {THORAX} {71}~({6}) ({2016}) {549--554}.

\bibitem{notter2000lung}
R.~Notter, Lung Surfactants: Basic Science and Clinical Applications, Lung
  Biology in Health and Disease, Taylor \& Francis, 2000.

\bibitem{Hohlfeld2002}
J.~Hohlfeld, {The role of surfactant in asthma}, {Respiratory Research}
  {3}~({1}).

\bibitem{Baritussio2004}
A.~Baritussio, {Lung surfactant, asthma, and allergens - A story in evolution},
  {American Journal of Respiratory and Critical Care Medicine} {169}~({5})
  ({2004}) {550--551}.

\bibitem{Wright2001}
T.~Wright, R.~Notter, Z.~Wang, A.~Harmsen, F.~Gigliotti, {Pulmonary
  inflammation disrupts surfactant function during Pneumocystis carinii
  pneumonia}, {Infection and Immunity} {69}~({2}) ({2001}) {758--764}.

\bibitem{Willson2008}
D.~F. Willson, P.~R. Chess, R.~H. Notter, {Surfactant for pediatric acute lung
  injury}, {Pediatric Clinics of North America} {55}~({3}) ({2008}) {545--575}.

\bibitem{Schwab2009}
U.~Schwab, K.~H. Rohde, Z.~Wang, P.~R. Chess, R.~H. Notter, D.~G. Russell,
  {Transcriptional responses of Mycobacterium tuberculosis to lung surfactant},
  {Microbial Pathogenesis} {46}~({4}) ({2009}) {185--193}.

\bibitem{Raghavendran2011}
K.~Raghavendran, D.~Willson, R.~N. Notter, {Surfactant Therapy for Acute Lung
  Injury and Acute Respiratory Distress Syndrome}, {Critical Care Clinics}
  {27}~({3}) ({2011}) {525--559}.

\bibitem{Chroneos2009}
Z.~C. Chroneos, et~al., Pulmonary surfactant and tuberculosis, Tuberculosis 89
  (2009) 10 -- 14.

\bibitem{Chimote2005}
G.~Chimote, R.~Banerjee, Lung surfactant dysfunction in tuberculosis: Effect of
  mycobacterial tubercular lipids on dipalmitoylphosphatidylcholine surface
  activity, Colloids and Surfaces B: Biointerfaces 45~(3-4) (2005) 215 -- 223.

\bibitem{Hasegawa2003}
T.~Hasegawa, R.~M. Leblanc, Aggregation properties of mycolic acid molecules in
  monolayer films: a comparative study of compounds from various acid-fast
  bacterial species, Biochimica et Biophysica Acta (BBA) - Biomembranes
  1617~(1-2) (2003) 89 -- 95.

\bibitem{Wang2008}
Z.~Wang, et~al., Peripheral cell wall lipids of mycobacterium tuberculosis are
  inhibitory to surfactant function, Tuberculosis 88~(3) (2008) 178 -- 186.

\bibitem{Grotberg2011}
J.~B. Grotberg, {Respiratory fluid mechanics}, {Physics of Fluids} {23}~({2}).

\bibitem{Halpern1993}
D.~Halpern, J.~Grotberg, {Surfactant Effects on Fluid-Elastic Instabilities of
  Liquid-lined Flexible Tubes - A Model of Airway-Closure}, {Journal of
  Biomechanical Engineering-Transactions of the ASME} {115}~({3}) ({1993})
  {271--277}.

\bibitem{Chimote2008}
G.~Chimote, R.~Banerjee, {Effect of mycobacterial lipids on surface properties
  of Curosurf (TM): Implications for lung surfactant dysfunction in
  tuberculosis}, {Respiratory Physiology \& Neurobiology} {162}~({1}) ({2008})
  {73--79}.

\end{thebibliography}

\end{document}